\documentclass[11pt,twoside,dvips]{article}
\usepackage{verbatim}
\usepackage{amssymb}
\usepackage{amsfonts}
\usepackage{amsmath}
\usepackage{pifont}
\usepackage[spanish,english]{babel}
\usepackage[pdftex]{graphicx}
\usepackage{verbatim}
\usepackage{epsfig}
\usepackage{bm}
\usepackage{longtable}
\usepackage{fancyhdr}
\begin{document}

\fancypagestyle{plain}{%
\fancyhf{}%
\fancyhead[LO, RE]{XXXVIII International Symposium on Physics in Collision, \\ Bogot\'a, Colombia, 11-15 september 2018}}

\fancyhead{}%
\fancyhead[LO, RE]{XXXVIII International Symposium on Physics in Collision, \\ Bogot\'a, Colombia, 11-15 september 2018}

\title{Five Non-Fritzsch Texture Zeros for Quarks Mass Matrices in the Standard 
Model}
\author{Yithsbey Giraldo$\thanks{%
e-mail: yithsbey@gmail.com}$, Eduardo Rojas$\thanks{%
e-mail: eduro4000@gmail.com}$, \\ Departamento de F\'isica, Universidad de 
Nari\~no, \\  A.A. 1175, San Juan de Pasto, Colombia
}
%EndAName
\date{}
\maketitle

\begin{abstract}
In the Standard Model, we obtain a non-Fritzsch
like configuration with five texture zeros for the
quark mass matrices. This matrix generates the
quark masses, the inner angles of the CKM unitary triangle, and the CP-violating 
phase in the
quark sector. This work can be applied to the
PMNS matrix in the lepton sector, by assuming
Dirac masses for the neutrinos, where non-trivial
predictions for the neutrino masses and mixing
angles are expected.
\end{abstract}

\section{Introduction}
In models such as the Standard Model~(SM) or its extensions,
where the 
right-handed fields are singlets under $SU (2)$,
it is  always 
possible  to choose a suitable basis for the right-handed quarks, 
so that the resultant quark mass matrices become 
hermitian~\cite{Giraldo:2015cpp,b30,b1,Ponce:2011qp}. 
Consequently, without loss of generality, the  mass matrices for the up- or down-type  quarks  can be written as:
$
 M_u^\dag=M_u\:\textrm{and}\: M_d^\dag=M_d.
$
Another consequence for models like the  SM where there is a 
freedom to make independent unitary transformations for left- and right-handed quarks, while keeping the gauge currents 
invariants, is the existence of infinite representations for the mass matrices,
all of them unitarily equivalent. 
In this analysis the same  ``weak-basis''~(WB) transformation must be applied 
 to $M_u$ and $M_d$  in order to keep the charged current invariant~\cite{b0,b30}, i.e., :
\(
 M_u\rightarrow M'_u=U^\dag M_uU,\; M_d\rightarrow M'_d=U^\dag M_d U,
\)
where $U$ is an arbitrary unitary matrix.
To ensure that the charged currents are invariant, it is enough to keep 
the Cabibbo-Kobayashi-Maskawa~CKM matrix invariant. 
%%%%%%%%%%%%%%%%%%%%%%%%%%%%%%%%%%%%%%%%%%%%%%%%%%%%%%%
Note that the observed parameters for the CKM mixing matrix 
 are fitted at the electroweak scale $\mu =m_Z$, so,
 it is necessary to use the quark masses (in MeV units) at the 
same scale~\cite{b6x,b18}:
{\small
\begin{equation}
\label{17a}
 m_u=1.38^{+0.42}_{-0.41}\,,\: m_c=638^{+43}_{-84},\: m_t=172100\pm{1200}\,,\\
m_d=2.82\pm0.48\,,\: m_s=57^{+18}_{-12}\,,\: m_b=2860^{+160}_{-60}.
\end{equation}}
The CKM mixing matrix~\cite{b6x} is a $3\times3$ unitary matrix,
 which can be parametrized by three mixing angles and the 
CP-violating phase. Usually it has the following standard 
choice
{\small
\begin{equation}
\label{3.2}
 V=\begin{pmatrix}
    V_{ud}&V_{us}&V_{ub}
\\
V_{cd}&V_{cs}&V_{cb}
\\
V_{td}&V_{ts}&V_{tb}
   \end{pmatrix}=\begin{pmatrix}
    c_{12}\,c_{13}&s_{12}\,c_{13}&s_{13}\,e^{-i\delta}\\
-s_{12}\,c_{23}-c_{12}\,s_{23}\,s_{13}\,e^{i\delta}& 
c_{12}\,c_{23}-s_{12}\,s_{23}\,s_{13}\,e^{i\delta}&
s_{23}\,c_{13}\\
s_{12}\,s_{23}-c_{12}\,c_{23}\,s_{13}\,e^{i\delta}&-c_{12}\,s_{23}-s_{12}\,c_{23
}\,s_{13}\,e^{i\delta}&c_{23}\,c_{13}
   \end{pmatrix},
\end{equation}}
where $s_{ij}=\sin\theta_{ij},c_{ij}=\cos\theta_{ij}$, and the angles are chosen to lie in 
the first quadrant, so $s_{ij},c_{ij}\ge0$. And $\delta$ is the phase responsible for all 
CP-violating phenomena in flavor-changing processes in the SM.
The Wolfenstein parametrization
exhibits the experimental hierarchy {\small $s_{13}\ll s_{23}\ll s_{12}\ll1$}. 
The fit results for the values of all nine CKM elements are:
$s_{12}= 0.22537 \pm0.00061$, 
$s_{23}=0.0413\pm0.0012$, $s_{13}e^{i\delta}=(0.00355\pm0.00016)e^{
i(1.250\pm0.056)}.$

%%%%%%%%%%%%%%%%%%%%%%%%%%%%%%%%%%%%%%%%%%%%%%%%%%%%%
%%%%%%%%%%%%%%%%%%%%%%%%%%%%%%%%%%%%%%%%%%%%%%%%%%%%%%%%%%%%%
\section{Five numeric texture zeros}
\label{sIII}
 By performing WB transformations on the quark mass matrices, 
 it is always possible to obtain one of the following  arrangements
~\cite{b26}:
{\small
\begin{equation}
\label{2.8}
 M_u=D_u=\begin{pmatrix}
          \lambda_{1u}&0&0\\
0&\lambda_{2u}&0\\
0&0&\lambda_{3u}
         \end{pmatrix},\:
 M_d=VD_dV^\dag,
\:\:\textrm{or},\:\:
 M_u=V^\dag D_uV,\:
 M_d=D_d=\begin{pmatrix}
          \lambda_{1d}&0&0\\
0&\lambda_{2d}&0\\
0&0&\lambda_{3d}
         \end{pmatrix}.
\end{equation}}
Therefore, because of their simplicity,
it is convenient  to consider them as the  
initial representations  for the quark mass matrices~\cite{b0,b3,b1,b26} ---where $V$ is 
the CKM mixing matrix, and the eigenvalues $|\lambda_{iq}|$ ($i=1,2,3$) are the 
up-~($q=u$) and down-~($q=d$) quark masses--- and 
{\small
\begin{equation}
\label{2.9}
 %\begin{split}
|\lambda_{1u}|=m_u, |\lambda_{2u}|=m_c, |\lambda_{3u}|=m_t,
|\lambda_{1d}|=m_d, |\lambda_{2d}|=m_s, 
|\lambda_{3d}|=m_b,\:\textrm{where}\: 
|\lambda_{1q}|\ll|\lambda_{2q}|\ll|\lambda_{3q}|.
%\end{split}
\end{equation}}
%

%%%%%%%%%%%%%%%%%%%%%%%%%%%%%%%%%%%%%%%

The properties of the WB transformations allow us to use the bases~(\ref{2.8}) 
as 
the initial matrices to generate any physical structure in the 
quark mass matrix sector. If there are texture zeros, this transformation can 
find them. Since some texture zeros are in the diagonal elements of the 
hermitian mass matrices, it implies that at least one and at most two of their 
eigenvalues are negative~\cite{b0}. Also, in the case of two negative 
eigenvalues, these mass matrices can be reduced to having one by adding a 
negative sign in the bases~(\ref{2.8}), as follows:
$
M_u=-(-M_u) \quad\textrm{or}\quad M_d=-(-M_d),
$
such that  WB transformations for terms in parentheses can be implemented. 
Therefore, 
without losing generality, the texture zeros in the models can be obtained by 
assuming that each quark mass matrix, $M_u$ and $M_d$, contains precisely one 
single negative eigenvalue~\cite{b1}, i.e.,

{%\small
\begin{equation}
 \label{2.12} \textrm{$\lambda_{iq}$ is negative for one value of $i$ and 
positive for the others.}
\end{equation}}

%%%%%%%%%%%%%%%%%%%%%%%%%%%%%%%%%%%%%%%%%%%%%%%%%%%%%
%%%%%%%%%%%%%%%%%%%%%%%%%%%%%%%%%%%%%%%%%%%%%%%%%%%%%

Each {\it realistic} quark mass matrix can contain, at most, three texture 
zeros. Also, there are only two possible patterns according to the distribution 
of the three zeros in the elements of the mass matrix. In the first case, the 
mass matrix has a single zero in the diagonal elements, while in the other 
case, 
it has only two zeros in the diagonal entries. The two respective basic 
patterns are as follows:
{\small
\begin{equation}
\label{3.1}
M_{1q}=\begin{pmatrix}
      0& |\xi_q|& 0\\
|\xi_q|& \gamma_q& 0\\
0& 0& \alpha_q
     \end{pmatrix}
 ,\quad
 M_{2q}=\begin{pmatrix}
      0& |\xi_q|& 0\\
|\xi_q|& 0& |\beta_q|\\
0& |\beta_q|& \alpha_q
     \end{pmatrix},
\end{equation}}
where $(q=u\;\textrm{or}\; d)$, and we can observe that by making WB 
transformations of the form $p_i\:M_{1,2q}\:p_i^T$ and considering all 
permutation matrices $p_i$
 {\tiny
$\begin{pmatrix}
      1&&\\
&&1\\
&1&
     \end{pmatrix},
   \begin{pmatrix}
     &&1 \\
&1& \\
1&& 
     \end{pmatrix},
    \begin{pmatrix}
     &1& \\
1&& \\
 &&1
     \end{pmatrix},
   \begin{pmatrix}
      &&1 \\
1&& \\
&1&
     \end{pmatrix},
\begin{pmatrix}
     &1& \\
&&1\\
1&& 
     \end{pmatrix},
$}
%\noindent
we get as many viable cases as possible for each pattern considered. All viable 
three-zero textures for quark mass 
matrices are summarized here. These patterns are general, and including phases 
is not necessary, as they can be absorbed by the other mass matrix~($u$ or $d$) 
through a WB transformation.
%
%%%%%%%%%%%%%%%%%%%%%%%%%%%%%%%%%%%%%%%%%%%%%5
%\subsection{Two-zero diagonal pattern}
%
Let us start with the standard representation of the pattern of two zeros in 
the 
diagonal entries, $M_{2q}$, expression~\eqref{3.1}, which coincides with the 
matrix~\eqref{30} by making  $\gamma_q=0$. Its diagonalization matrix, 
$U_{2q}$, satisfies Eq.~\eqref{31x}, from which the results~\eqref{3.18} are 
derived; so, we have the following:
{\small
\begin{equation}
\label{83x}
\alpha_q=\lambda_{1q}+\lambda_{2q}+\lambda_{3q},\:\:
|\xi_q|=\sqrt{\frac{-\lambda_{1q}\lambda_{2q}\lambda_{3q}}{\alpha_q}},\:\:
%\label{84}
|\beta_q|=\sqrt{-\frac{(\lambda_{1{q}}+\lambda_{2{q}})(\lambda_{1{q}}+\lambda_{
3{q}})(\lambda_{2{q}}+
\lambda_{3{q}})}{\alpha_q}}.
\end{equation}}
The result~(\ref{83x}) for $|\xi_q|$ must be a real number, and because only an 
eigenvalue $\lambda_{iq}$ is assumed negative, Eq.~\eqref{2.12}, we have that $ 
\alpha_q>0$, where, together with~(\ref{83x}) for $\beta_q$,  and 
hierarchy~(\ref{2.9}), only 
one possibility is allowed:
{%\small
\begin{equation}
\label{4.10}
\lambda_{1{q}},\lambda_{3{q}}>0\quad\textrm{and}\quad\lambda_{2{q}}<0,
\quad\textrm{with}\quad \alpha_q>0.
\end{equation}}
According to~\eqref{32x}, the matrix that diagonalizes $M_{2q}$ is
{\small
%\begin{widetext}
\begin{equation} 
\label{32}
{%\Large
 U_{2q}=\begin{pmatrix}
     e^{ix} 
\sqrt{\frac{\lambda_{2q}\lambda_{3q}(\alpha_q-\lambda_{1q})}{\alpha_q(\lambda_{
2q}-
\lambda_{1q})(\lambda_{3q}-\lambda_{1q})}}&-e^{iy}
\sqrt{\frac{\lambda_{1q}\lambda_{3q}(\lambda_{2q}-\alpha_q)}{\alpha_q(\lambda_{
2q}-\lambda_{1q})
(\lambda_{3q}-\lambda_{2q})}}&
\sqrt{\frac{\lambda_{1q}\lambda_{2q}(\alpha_q-\lambda_{3q})}{\alpha_q(\lambda_{
3q}-\lambda_{1q})(\lambda_{3q}-\lambda_{2q})}}\\
&&&\\[-2mm]
e^{ix}\sqrt{\frac{\lambda_{1q}(\lambda_{1q}-\alpha_q)}{(\lambda_{2q}-\lambda_{1q
})(\lambda_{3q}-
\lambda_{1q})}}&
e^{iy}\sqrt{\frac{\lambda_{2q}(\alpha_q-\lambda_{2q})}{(\lambda_{2q}-\lambda_{1q
})(\lambda_{3q}-
\lambda_{2q})}}&
\sqrt{\frac{\lambda_{3q}(\lambda_{3q}-\alpha_q)}{(\lambda_{3q}-\lambda_{1q}
)(\lambda_{3q}-
\lambda_{2q})}}\\
&&&\\[-2mm]
-e^{ix}\sqrt{\frac{\lambda_{1q}(\alpha_q-\lambda_{2q})(\alpha_q-\lambda_{3q})}{
\alpha_q
(\lambda_{2q}-\lambda_{1q})(\lambda_{3q}-\lambda_{1q})}}&
-e^{iy}\sqrt{\frac{\lambda_{2q}(\alpha_q-\lambda_{1q})(\lambda_{3q}-\alpha_q)}{
\alpha_q
(\lambda_{2q}-\lambda_{1q})(\lambda_{3q}-\lambda_{2q})}}&
\sqrt{\frac{\lambda_{3q}(\alpha_q-\lambda_{1q})(\alpha_q-\lambda_{2q})}{\alpha_q
(\lambda_{3q}-\lambda_{1q})(\lambda_{3q}-\lambda_{2q})}}
     \end{pmatrix},} 
\end{equation}}
%\end{widetext}}
\noindent
where $\alpha_q$ is in Eq.~\eqref{83x}.
%
%%%%%%%%%%%%%%%%%%%%%%%%%
%\subsubsection{The diagonal-down basis}
%
Performing a WB transformation on the second base of~\eqref{2.8}, using, in 
this case, the 
unitary matrix given in~\eqref{32} with $q=d$ (i.e., $U_{2d}$), we have
{\small
\begin{equation}
\label{3.6z}
M_{d}^{\prime}=U_{2d}(D_d)U_{2d}^\dag=\begin{pmatrix}
      0& |\xi_{d}|& 0\\
|\xi_{d}|& 0& |\beta_{d}|\\
0& |\beta_{d}|& \alpha_{d}
     \end{pmatrix},\:\:
%\label{3.6y}
M_{u}^\prime=U_{2d}\,(V^\dag D_uV)\,U_{2d}^\dag\,,
\end{equation}}
\noindent
where the equation~\eqref{31x} was taken into account. According 
to~\eqref{4.10}, in this case
$
\lambda_{1{d}},\: \lambda_{3{d}}>0,$ $\lambda_{2{d}}<0$, and $ 
\alpha_{d}=\lambda_{1d}+
\lambda_{2d}+\lambda_{3d}>0.
$
The calculations are simplified if we define the following variables for the 
phases introduced in~\eqref{32}:
\begin{align}
\label{89}
 e^{ix}=
x_1+ix_2, \quad
e^{iy}=
y_1+iy_2,
\end{align}
where $x_1^2+x_2^2=1$ and $y_1^2+y_2^2=1$, and it is satisfied that
$
\label{4.18}
 |x_1|,|x_2|,  |y_1|,|y_2|\leq1.
$
With these definitions, and using experimental data~\eqref{17a} 
and~\eqref{3.2}, 
the elements of the  matrix $M_{u}^\prime$, in~(\ref{3.6z}) become 
{\it surfaces} for the points $(x_1,x_2,y_1,y_2)$ in~$\mathbb{R}^4$ 
for each case considered:  $\lambda_{1u}=-m_u$ or 
$\lambda_{2u}=-m_c$ or $\lambda_{3u}=-m_t$. Taking~\eqref{89} into account, the 
analysis of these surfaces shows that only elements (1,2) and (1,3) of 
$M_{u}^\prime$ can give solutions equal to zero (texture zeros).
Let's take the case $\lambda_{1u}=-m_u$ as an example; we obtain the best 
results considering the following masses for quarks (in MeV units): 
$m_u=1.7160$, 
$m_d=2.9042$, $m_s=65.0$, $m_c=567.0$, $m_b=2860.0$, $m_t=172100$, which are 
close to 
the central values and are within the range allowed by~\eqref{17a}. The 
solutions are: 
$
 x_1 = 0.68499$, $y_1 =-0.50043$,
$x_2 =0.72855$, $y_2 = -0.86578$,
with the component $M_u^\prime(1,1)=0$. The corresponding numerical matrices 
obtained for the quark masses with five texture zeros are the following~\cite{Giraldo:2015cpp}
%
%\begin{widetext}
{\scriptsize	
\begin{subequations}
\label{4.22}
\begin{align}
 M_u^{\prime}&=\begin{pmatrix}
     0&0& -79.323 + 154.72 i
\\
 0& 5539.2& 28126 + 6112.8 i
\\
 -79.323 - 154.72 i&28126 - 6112.8 i& 167130
     \end{pmatrix}\text{MeV},\\
M_d^{\prime}&=\begin{pmatrix}
    0&13.891& 0
\\
 13.891& 0& 421.41 
\\
 0& 421.41 &2797.9
    \end{pmatrix}\text{MeV},
\end{align}
\end{subequations}
}
and their diagonalization matrices are respectively
{\scriptsize
\begin{align*}
 U'_u&=\begin{pmatrix}
  0.67626 + 0.73481 i& -0.050244 + 0.013892i& 
-0.00044950
+ 0.00088826i
\\
 0.027505 - 0.043184 i& -0.49704 - 0.84931 i&
  0.16658 + 0.035285 i
\\
 -0.0034086 + 0.0092482 i& 0.11507 + 0.12514 i&
 0.98538 - 0.00519999 i
     \end{pmatrix},\\
U'_d&=\begin{pmatrix}
0.67018 + 0.71279 i& 0.10352 + 0.17909 i& 0.00070804
\\
 0.14011 + 0.14902 i& -0.48439 - 0.83802 i& 0.14578
\\
 -0.021126 - 0.022469 i& 0.071301 + 0.12335 i& 0.98932
    \end{pmatrix},
\end{align*}
}
%\end{widetext}
%
 which gives the correct CKM mixing matrix with a precision level 
of 1$\sigma$~\cite{b43,Verma:2015mgd,Verma:2017ppl}, $V=U_u^{\prime\dag} U'_d$, 
including the phase 
responsible for the CP violation. The case $\lambda_{2u}=-m_c$  has already 
been done in article~\cite{b1}. 
We also note that the first  diagonal base~\eqref{2.8} and the pattern with a 
zero in the diagonal~\eqref{3.1} do not give additional consistent solutions 
with five texture zeros.

%%%%%%%%%%%%%%%%%%%%%%%%%%%%%%%%%%%%%%%%%%%%
%%%%%%%%%%%%%%%%%%%%%%%%%%%%%%%%%%%%%%%%%%%%
\section{Conclusions}
\label{sV}

We have made a complete study of the texture zeros in the quark sector of the 
SM, starting from general quark mass matrices, based on the WB transformation 
property~\cite{b1}, to generate any possible mass matrix configuration. This 
result allowed us to use specific basis, \eqref{2.8}, to reproduce 
as 
many texture zeros as possible. In this way, we discovered a numerical texture 
pattern consisting of five zeros, including permutations, whose matrix 
representation is in~\eqref{4.22}; 
the pattern is not Fritzsch type~\cite{b30} because of the way texture zeros 
are 
present. It suggests the 
construction of a model with nine parameters that involves mass and mixing 
relations, while, at the same time, gives precise masses for quarks, CKM mixing 
angles, and the phase responsible for the CP violation at a confidence level.
There are no additional representations of five texture zeros apart 
from the one given in article~\cite{b1}. By assuming Dirac masses for the
neutrinos, next step is to apply our results to the
PMNS matrix in the lepton sector where non-trivial results are 
expected~\cite{Hernandez:2014zsa,Giraldo19}.

%%%%%%%%%%%%%%%%%%%%%%%%%%%%%%%%%%%%%%%%%%%%%%%%%%5
%%%%%%%%%%%%%%%%%%%%%%%%%%%%%%%%%%%%%%%%%%%%%%%%%%5
%%%%%%%%%%%%%%%%%%%%%%%%%%%%%%%%%%%%%%%%%%%%%%%%%%5

\section*{Acknowledgment}
We thank R.~H. Benavides, L. Mu\~noz and W.~ A. Ponce for their collaboration on these subjects.
This work was partially supported by VIPRI in the Universidad de
Nari\~no, Approval Contract No. 024.
\appendix

\section{Matrix with two texture zeros}
\label{sIIIC}
Consider the following structure for the up-~($q=u$) and down-~($q=d$) quark 
mass matrix:
{\small
\begin{equation}
\label{30}
 M_q=\begin{pmatrix}
      0&|\xi_{q}|&0\\
|\xi_{q}|&\gamma_{q}&|\beta_{q}|\\
0&|\beta_{q}|&\alpha_{q}
     \end{pmatrix}.
\end{equation}}
Owing to the hermiticity of $M_q$, $\gamma_{q}$ and $\alpha_{q}$ are real 
numbers. The phases of the parameters outside the diagonal can be included 
later 
using a WB transformation. By diagonalizing the mass matrix $M_q$, we have
{\small
\begin{equation}
\label{31x}
U_q^\dag M_qU_q=D_q,
\end{equation}}
where the $\lambda_{iq}$~$(i=1,2, 3)$ are defined in~\eqref{2.9} and $D_q$ 
in~\eqref{2.8}. The 
parameters $\gamma_{q}$, $|\xi_{q}|$ and $|\beta_{q}|$  can be expressed in 
terms of $\lambda_{iq}$ and $\alpha_q$. For this, we apply the invariant matrix 
functions $\text{tr} M_q$, $\textrm{tr} M_q^2 $ and $\det M_q$, 
on~\eqref{31x}. Results
{\small
\begin{equation}
\label{3.18}
%\begin{split}
\gamma_q=\lambda_{1q}+\lambda_{2q}+\lambda_{3q}-\alpha_q,\:\:
|\xi_q|=\sqrt{\frac{-\lambda_{1q}\lambda_{2q}\lambda_{3q}}{\alpha_q}},\:\:
|\beta_q|=\sqrt{\frac{(\alpha_q-\lambda_{1q})(\alpha_q-\lambda_{2q})(\lambda_{
3q}-\alpha_q)}{\alpha_q}}.
%\end{split}
\end{equation}}
The expressions~\eqref{3.18} are real, so the parameter $\alpha_q$ is confined 
to one interval. Let's see the different possibilities: If $\lambda_{1q}<0$, 
$\lambda_{2q}>0$ and $\lambda_{3q}>0$ then $ 
|\lambda_{2q}|<\alpha_q<|\lambda_{3q}|$, if $\lambda_{1q}>0$, $\lambda_{2q}<0$ 
and $\lambda_{3q}>0$ then $|\lambda_{1q}|<\alpha_q<|\lambda_{3q}|$, if 
$\lambda_{1q}>0$, $\lambda_{2q}>0$ and $\lambda_{3q}<0$ then 
$|\lambda_{1q}|<\alpha_q<|\lambda_{2q}|.$
In the previous analysis, we take into account the hierarchy~(\ref{2.9}) and 
the assumption~\eqref{2.12}. The exact analytical result for the matrix~($U_q$)
that diagonalizes to $M_q$ in~\eqref{30} is~\cite{b30, b1,Ponce:2013nsa}
%
%\begin{widetext}
{\scriptsize
\begin{equation} 
\label{32x}
{
 U_q=\begin{pmatrix}
     e^{ix} 
\frac{|\lambda_{3q}|}{\lambda_{3q}}\sqrt{\frac{\lambda_{2q}\lambda_{3q}
(\alpha_q-\lambda_{1q})}{\alpha_q(\lambda_{2q}-\lambda_{1q})(\lambda_{3q}
-\lambda_{1q})}}&e^{iy}\frac{|\lambda_{2q}|}{\lambda_{2q}}
\sqrt{\frac{\lambda_{1q}\lambda_{3q}(\lambda_{2q}-\alpha_q)}{\alpha_q(\lambda_{
2q}-\lambda_{1q})
(\lambda_{3q}-\lambda_{2q})}}&
\sqrt{\frac{\lambda_{1q}\lambda_{2q}(\alpha_q-\lambda_{3q})}{\alpha_q(\lambda_{
3q}-\lambda_{1q})
(\lambda_{3q}-\lambda_{2q})}}\\
&&&\\[-2mm]
-e^{ix}\frac{|\lambda_{2q}|}{\lambda_{2q}}\sqrt{\frac{\lambda_{1q}(\lambda_{1q}
-\alpha_q)}
{(\lambda_{2q}-\lambda_{1q})(\lambda_{3q}-\lambda_{1q})}}&
e^{iy}\sqrt{\frac{\lambda_{2q}(\alpha_q-\lambda_{2q})}{(\lambda_{2q}-\lambda_{1q
})
(\lambda_{3q}-\lambda_{2q})}}&
\frac{|\lambda_{3q}|}{\lambda_{3q}}\sqrt{\frac{\lambda_{3q}(\lambda_{3q}
-\alpha_q)}
{(\lambda_{3q}-\lambda_{1q})(\lambda_{3q}-\lambda_{2q})}}\\
&&&\\[-2mm]
e^{ix}\frac{|\lambda_{2q}|}{\lambda_{2q}}\sqrt{\frac{\lambda_{1q}
(\alpha_q-\lambda_{2q})
(\alpha_q-\lambda_{3q})}{\alpha_q(\lambda_{2q}-\lambda_{1q})(\lambda_{3q}
-\lambda_{1q})}}&
-e^{iy}\frac{|\lambda_{3q}|}{\lambda_{3q}}\sqrt{\frac{\lambda_{2q}
(\alpha_q-\lambda_{1q})
(\lambda_{3q}-\alpha_q)}{\alpha_q(\lambda_{2q}-\lambda_{1q})(\lambda_{3q}
-\lambda_{2q})}}&
\sqrt{\frac{\lambda_{3q}(\alpha_q-\lambda_{1q})(\alpha_q-\lambda_{2q})}
{\alpha_q(\lambda_{3q}-\lambda_{1q})(\lambda_{3q}-\lambda_{2q})}}
     \end{pmatrix},}
\end{equation}}
%\end{widetext}
%
where we have added phases to make the CKM matrix compatible with the 
convention 
chosen in~\eqref{3.2}, something that was justified in~\cite{b26}. 

\bibliographystyle{unsrt}
\bibliography{iopart-num}

\end{document}